\documentclass[pra,a4paper]{revtex4}

\usepackage{graphicx}
\usepackage{bbm}
\usepackage{amsmath}
\usepackage{amsfonts}
\usepackage{amssymb}

\begin{document}

\title{The Dynamical Nonlocality of Neutral Kaons and the Kaonic Quantum Eraser}

%
\author{Beatrix C. Hiesmayr}
\email{Beatrix.Hiesmayr@univie.ac.at}
\affiliation{Research Center for Quantum Information, Institute of Physics,
Slovak Academy of Sciences, Dúbravská cesta 9, 84511 Bratislava, Slovakia;\\ University of Vienna, Faculty of Physics, Boltzmanngasse 5, 1090 Vienna, Austria.
}

\begin{abstract}
Testing quantum foundations for systems in high
energy physics gets currently more and more attention e.g. witnessed for entangled neutral K-mesons by the approved programme of the KLOE collaboration
at the accelerator facility DAPHNE (Frascati, Italy). We focus on this quantum system in high energy physics and discuss two topics, Bell inequalities and the kaonic quantum eraser, and show how the neutral kaon system differs from systems of ordinary matter and light. In detail, we show a relation of the imbalance of matter and antimatter to the violation of a Bell inequality and discuss another Bell inequality which is maximally violated for a non-maximally entangled state though neutral kaons can be considered as two state systems. We compare in general this system in high energy physics with bipartite qudits. Last but not least  we review the quantum marking and eraser procedure and explain why neutral kaons offer more eraser possibilities than usual quantum systems.
\end{abstract}

\maketitle


\section{Introduction}

Particle physics has become an interesting testing ground for
questions of foundations of quantum mechanics as the accelerator
experiments reach higher and higher precisions. This allows not only for deep
tests of the standard model and possibly beyond but as well on
foundations of quantum mechanics. E.g. massive meson--antimeson systems entangled in their flavour are specially suitable to test foundations. In this paper we focus at the neutral kaon system entangled in strangeness and discuss Bell inequalities and
the quantum marking and eraser procedures which are one of the three issues of the recently approved programme of the KLOE collaboration (see Ref.~\cite{Hiesmayr11}) at the DAPHNE accelerator in Frascati (Italy). In detail we
\begin{itemize}
\item show a Bell inequality which is related to a symmetry violation in
particle physics, i.e. connecting two different concepts in physics \cite{Hiesmayr4, Hiesmayr3}
\item show a Bell inequality which can be tested directly in experiments,
but is not violated for the maximally entangled spin singlet state \cite{Hiesmayr3},
but surprisingly for a non--maximally entangled
state~\cite{Hiesmayr1}. This opens --- for the first time --- the
possibility of a direct experimental test whether local realistic
theories can be refuted
\item discuss the generalized Bell inequality for any pure entangled state
\item discuss quantum marking and erasure experiments with
neutral kaons~\cite{Hiesmayr7,Hiesmayr8} which offers two more
erasure options not available by other quantum systems
\end{itemize}
Of course also other features can be/have been discussed, e.g. one can discuss decoherence models for entangled K--mesons  and B-mesons (e.g. Refs.~\cite{Hiesmayr6,HiesmayrDeco}), its connection to entanglement measures and show the
recent bounds obtained by various accelerator experiments  (for kaons see e.g. Refs.~\cite{Ambrosio,DiDomenico,CPLEAR} or for B-mesons see e.g. Ref.~\cite{Go}).
In a quantum gravity picture space--time might be subjected to inherent fluctuations at the Planck scale. This would cause pure states to evolve to mixed states, i.e. decoherence of apparently isolated matter systems would occur suggested by Hawking~\cite{Hawking}. This decoherence necessarily implies that the quantum mechanical operator generating CPT transformations cannot be consistently defined~\cite{Wald}. Such a model was formulated in Ref.~\cite{Ellis,Mavromatos} and estimates of the parameter by the KLOE experiment can be found in Ref.~\cite{DiDomenicoCPT}. An information theoretic analyzes of the neutral kaon
system~\cite{Hiesmayr9,Hiesmayr10} is also available offering a comparison to usual quantum systems.

In summary, understanding the nature of entanglement and its
manifestation needs also studies of entanglement at different energy
scales and as recent works (e.g. about high--dimensional genuine
multipartite entanglement~\cite{Hiesmayr12,Hiesmayr13a,Schauer}) show there are still many open questions
concerning entanglement.

\section{Bell inequalities for neutral kaons}

The neutral K-mesons or simply kaons are bound states of quarks and
anti--quarks or more precis the strangeness state $+1$, $K^0$, is
composed of an anti--strange quark and a down quark and the
strangeness state, $\bar K^0$, is composed of a strange and
anti--down quark. In accelerator experiments one can produce a spin
singlet state, i.e. one has the same scenario as Einstein, Podolsky
and Rosen considered in 1935 which in the following is written down for
different quantum systems (spin--$\frac{1}{2}$, ground/excited
state, polarisation entangled photons, single neutrons in an
interferometer, molecules, K--meson, B--mesons):
\begin{eqnarray}\label{antisymmetricBellstate}
|\psi^-\rangle&=&\frac{1}{\sqrt{2}}\big\lbrace
|\Uparrow\rangle_l\otimes|\Downarrow\rangle_r-|\Downarrow\rangle_l\otimes|\Uparrow\rangle_r\big\rbrace\nonumber\\
&=&\frac{1}{\sqrt{2}}\big\lbrace
|0\rangle_l\otimes|1\rangle_r-|1\rangle_l\otimes|0\rangle_r\big\rbrace\nonumber\\
&=&\frac{1}{\sqrt{2}}\big\lbrace
|H\rangle_l\otimes|V\rangle_r-|V\rangle_l\otimes|H\rangle_r\big\rbrace\nonumber\\
&=&\frac{1}{\sqrt{2}}\big\lbrace|I\rangle_l\otimes|\Uparrow\rangle_r-|II\rangle_l\otimes|\Downarrow\rangle_r\big\rbrace\nonumber\\
&=&\frac{1}{\sqrt{2}}\big\lbrace
|\textrm{early}\rangle_l\otimes|\textrm{late}\rangle_r-|\textrm{late}\rangle_l\otimes|\textrm{early}\rangle_r\big\rbrace\nonumber\\
&=&\frac{1}{\sqrt{2}}\big\lbrace
|K^0\rangle_l\otimes|\bar K^0\rangle_r-|\bar K^0\rangle_l\otimes|K^0\rangle_r\big\rbrace\nonumber\\
&=&\frac{1}{\sqrt{2}}\big\lbrace
|B^0\rangle_l\otimes|\bar B^0\rangle_r-|\bar B^0\rangle_l\otimes|B^0\rangle_r\big\rbrace\nonumber\\
&=&\dots\;.
\end{eqnarray}
Most Bell tests have been performed with polarisation entangled photons, e.g. under Einstein's locality condition \cite{Weihs}. A serious of experiments testing the entanglement of molecules has recently been performed, see e.g. Ref.~\cite{Hornberger}. For single neutrons experiments exist which entangle the outer degree of freedom (path in the interferometer) with inner degrees of freedom (spin), a violation of Bell's inequality can be found, even including the Berry phase, a geometrical phase. The Bell inequality can be violated if the observables are accordingly compensated for the phase \cite{HiesmayrBerry} which was recently experimentally verified~\cite{Sponar}. Different to systems of ordinary matter and light neutral kaons oscillate in time (strangeness
oscillation) and are decaying and therefore entanglement manifests itself considerably differently.

Analogous to entangled photon systems Bell
inequalities can be derived for neutral kaons, i.e. the most general Bell inequality
of the CHSH--type is given by (see Ref.~\cite{Hiesmayr3})
\begin{eqnarray}\label{chsh}
\lefteqn{S_{k_n,k_m,k_{n'},k_{m'}}(t_n,t_m,t_{n'},t_{m'})=}\nonumber\\
&&\left|
E_{k_n,k_m}(t_n,t_m)-E_{k_n,k_{m'}}(t_n,t_{m'})\right|+|E_{k_{n'},k_{m}}(t_{n'},t_m)+E_{k_{n'},k_{m'}}(t_{n'},t_{m'})|\leq
2\;.
\end{eqnarray}
The upper limit is obtained under the usual assumption for any local realistic theories, i.e. that the most general form of an expectation value is given by
\begin{eqnarray}
E_{k_n,k_m}(t_1,t_2)&=& \int d\lambda\; \rho(\lambda) O(k_n,t_n,\lambda)\cdot O(k_m,t_m,\lambda)\quad\textrm{with}\quad\int d\lambda\; \rho(\lambda)=1\;.
\end{eqnarray}
Here Alice chooses to measure on the kaon propagating to her left hand side
whether it is in the ``quasi--spin'' $|k_n\rangle$ or not, where $|k_n\rangle=\alpha |K^0\rangle+\beta |\bar K^0\rangle$,
and how long the kaon propagates, i.e. the time $t_n$ (corresponding to the distance from the source). The same options are
given to Bob for the kaon propagating to the right hand side. As in
the usual photon setup, Alice and Bob can choose among two settings.
We notice already that in the neutral kaon case we have more options
than in the photon case, we can vary in the quasi--spin space or
vary the detection times or both.

\subsection{An open quantum formalism describing neutral kaons}\label{decay}

In order to discuss fundamental questions of quantum mechanics we have to have an
appropriate model to describe the time evolution of the meson
system, i.e., oscillation ($K^0\leftrightarrow\bar K^0$) and decay.
In Ref.~\cite{HiesmayrOpen} it has been shown that a decaying
system can be handled with the open quantum system formulation, i.e.
a master equation of the Lindblad type. For that one has to enlarge
the original Hilbert space. The particle decay is then incorporated
by a certain Lindblad generator. With this master equation the
Hamiltonian of the whole system (``surviving'' and ``decaying''
states) is then Hermitian and trace preserving just like common
quantum systems. Moreover, the time evolution is clearly completely
positive and Markovian. The effective ``environment'' causing the
decay of the mesons is what in quantum field theory is the
quantum--chromo--dynamic vacuum.

The neutral kaon system is a decaying two--state system due to the
strangeness oscillation in time, $K^0\leftrightarrow \bar K^0$, and
is usually described via an effective Schr\"odinger equation which
we write in the Lioville von Neumann form ($\hbar\equiv1$)
\begin{eqnarray}\label{effectiveSchroedi}
\frac{d}{dt} \mathbf{\rho}&=&-i\, H_{eff}\; \mathbf{\rho}+i\,
\mathbf{\rho}\; H_{eff}^\dagger
\end{eqnarray}
where $\rho$ is a $2\times 2$ density matrix and $H_{eff}$ is
non-Hermitian. Using the usual Wigner-Weisskopf-approximation the
effective Hamilton can be decomposed as
$H_{eff}=H-\frac{i}{2}\Gamma$ where the mass matrix $H$ and the
decay matrix $\Gamma$ are both Hermitian and positive. To obtain
this Hamiltonian the weak interaction Hamilton is treated as a
perturbation and interaction between the final states is neglected,
in particular the decay states of the neutral kaons, e.g. pions, are
assumed to be stable. The second approximation is to consider only
the first pole contribution and this leads to the exponential time
evolution of the two diagonal states of $H_{eff}$
\begin{eqnarray}\label{exptime}
|K_{S/L}(t)\rangle &=& e^{-i m_{S/L} t} e^{-\frac{\Gamma_{S/L}}{2}
t} |K_{S/L}\rangle\,,
\end{eqnarray}
$m_{S/L}$ and $\Gamma_{S/L}$ are the masses or decay constants of
the short or long lived kaons. What makes the neutral kaon systems
so attractive for many physical applications is the huge factor
between the two decay rates, i.e. $\Gamma_S\approx 600 \Gamma_L$.

Considering Eq.~(\ref{exptime})---which describes the surviving
components of the neutral kaon evolving in time---we notice that the
state is not normalized for $t>0$. Indeed, we are not including all
information on the system under investigation. For $t> 0$ the
neutral kaon is a superposition of the surviving components and
decaying components. In Ref.~\cite{HiesmayrOpen} the authors showed that by
at least doubling the original two--dimensional Hilbert space the
decay can be incorporated into the dissipator of the enlarged space
via a Lindblad operator. Let us start with the well known master
equation in the Lindblad form
\begin{eqnarray}\label{masterequation}
\frac{d}{dt} \rho&=&-i [\cal{H},\rho]-\cal{D}[\rho]
\end{eqnarray}
where the dissipator has the general form
\begin{eqnarray*}
\cal{D}[\rho]&=&\frac{1}{2}\sum_{j=0} ({\cal A}_j^\dagger{\cal
A}_j\rho+\rho{\cal A}_j^\dagger{\cal A}_j-2 {\cal A}_j\rho{\cal
A}_j^\dagger)\;.
\end{eqnarray*}

The density matrix is $4$ dimensional and lives on
$\textbf{H}_{tot}=\textbf{H}_s+\textbf{H}_f$ where $s/f$
denotes ``surviving'' and ``decaying'' or ``final'', respectively,
and has the following components
\begin{eqnarray*}
\rho&=&\left(\begin{array}{cc} \rho_{ss}&\rho_{sf}\\
\rho_{fs}&\rho_{ff}\end{array}\right)
\end{eqnarray*}
where $\rho_{ij}$ with $i,j=s,f$ denote $2\times 2$ matrices and
$\rho_{sf}=\rho_{fs}^\dagger$. The Hamiltonian $\cal{H}$ is the
Hamiltonian $H$ from the effective Hamiltonian $H_{eff}$ extended to
the total Hilbert space
\begin{eqnarray*}
{\cal H}=\left(\begin{array}{cc} H&0\\
0&0\end{array}\right)\, .
\end{eqnarray*}
To incorporate the decay ability we define one Lindblad generator,
e.g. $A_0$, to be
\begin{eqnarray*}
{\cal A}_0=\left(\begin{array}{cc} 0&0\\
B&0\end{array}\right)\qquad\textrm{with}\quad B:
\textbf{H}_s\rightarrow \textbf{H}_f\;,
\end{eqnarray*}
where $B^\dagger B=\Gamma$ and $\Gamma$ is the decay matrix of
$H_{eff}$. All other Lindblad generators ($j>0$) can only act on the
surviving component of the density matrix, i.e.:
\begin{eqnarray*}
{\cal A}_j=\left(\begin{array}{cc} A_j&0\\
0&0\end{array}\right)\qquad\textrm{with}\quad j\not=0\,.
\end{eqnarray*}

Rewritten the master equation for the density matrix on the total
Hilbert space $\textbf{H}_{tot}$ decomposed into the components of
$\rho$ we have
\begin{eqnarray*}
\dot{\rho}_{ss}&=&-i[H,\rho_{ss}]-\frac{1}{2}\lbrace
\underbrace{B^\dagger
B}_{\Gamma},\rho_{ss}\rbrace\,-\tilde D[\rho_{ss}],\\
\dot{\rho}_{sf}&=&-i H \rho_{sf}-\frac{1}{2} \underbrace{B^\dagger
B}_\Gamma\rho_{sf}-\frac{1}{2} \sum_jA_j^\dagger
A_j \rho_{sf}\;,\\
\label{rhoff} \dot{\rho}_{ff}&=&B\;\rho_{ss}\;B^\dagger\,,
\end{eqnarray*}
where $\tilde D[\rho_{ss}]=\frac{1}{2}\sum_{j=1} ( A_j^\dagger
A_j\rho+\rho A_j^\dagger A_j-2 A_j \rho A_j^\dagger)$ describes any
decoherence or dissipation which may occur. Indeed, we immediately see that this master equation describes the
original effective Schr\"odinger equation (\ref{effectiveSchroedi})
and thus the decay of neutral kaons (and in addition decoherence):
\begin{enumerate}
\item[(1)] By construction the time evolution of $\rho_{ss}$ is
independent of $\rho_{sf}, \rho_{fs}$ and $\rho_{ff}$.
\item[(2)] Furthermore $\rho_{sf},\rho_{fs}$ completely decouples from
$\rho_{ss}$. Consequently, these components $\rho_{sf}, \rho_{fs}$
can without loss of generality be chosen to be zero, they are not
physical and cannot be measured.
\item[(3)] With the initial condition $\rho_{ff}(0)=0$ the time
evolution is solely determined by $\rho_{ss}$ and simply given by
\begin{eqnarray}
\rho_{ff}(t)&=& B\int_{0}^{t}dt' \rho_{ss}(t')B^\dagger\,.
\end{eqnarray}
\end{enumerate}
And indeed, this are the properties which we expect to describe a
particle decay!

The extension for entangled kaons is straight forward given by replacing the Hamiltonian and decay generator, respectively
\begin{eqnarray}
\cal{H}&\longrightarrow& \cal{H}\otimes\mathbbm{1}+\mathbbm{1}\otimes\cal{H}\nonumber\\
A_0&\longrightarrow&A_0\otimes\mathbbm{1}+\mathbbm{1}\otimes A_0
\end{eqnarray}

Summarizing, we showed that particle decay and
decoherence/dissipation are related phenomena and that the Wigner
Weisskopf approximation is Markovian and completely positive.

\subsection{A Bell inequality sensitive to the imbalance of matter and antimatter in our universe}

Let us first choose all times equal zero and choose the quasi--spin
states $k_n=K_S, k_m=\bar K^0, k_{n'}=k_{m'}=K_1^0$ where $K_S$ is
the short--lived eigenstate, one eigenstate of the time evolution,
and $K_1^0$ is the ${\cal CP}$ plus eigenstate.  Here $\cal{C}$
stands for charge conjugation and  $\cal{P}$ for parity. The neutral
kaon is famous for violating the combined transformation ${\cal CP}$
for which Fitch and Turlay got the Nobel Prize in 1980. In
Ref.~\cite{Hiesmayr4} the authors showed that after optimizing and a further trick the
Bell inequality (\ref{chsh}) can be turned into
\begin{eqnarray}\label{BICP}
\delta\leq 0\;
\end{eqnarray}
where $\delta$ is the ${\cal CP}$ violating parameter in mixing.
Experimentally, $\delta$ corresponds to the leptonic asymmetry of
kaon decays which is measured to be $\delta=(3.27\pm0.12)\cdot
10^{-3}$. This value is in clear contradiction to the value required
by the BI above, i.e. by the premises of local realistic theories!
In this sense the violation of a symmetry in high energy physics is
connected to the violation of a Bell inequality, i.e. to
nonlocality. This is clearly not available for photons, they do not
violate the $\cal{CP}$ symmetry. Although the BI~(\ref{BICP}) is as loophole free as possible, the
probabilities or expectations values involved are not directly
measurable, because experimentally there is no way to distinguish
the short--lived state $K_S$ from the ${\cal CP}$ plus state $K_1^0$
directly.

\subsection{A Bell inequality sensitive to the strangeness oscillation}

Another choice for the generalized Bell
inequality (\ref{chsh}) is obtained by setting all quasispins equal $\bar K^0$ and varying all times
\begin{eqnarray}\label{BIstrangeness}
\lefteqn{S_{\bar K^0,\bar K^0,\bar K^0,\bar
K^0}(t_1,t_2,t_3,t_4)=}\nonumber\\&&| E_{\bar K^0,\bar K^0}(t_1,t_2)-E_{\bar K^0,\bar
K^0}(t_1,t_3)|+|E_{\bar K^0,\bar K^0}(t_4,t_2)+E_{\bar K^0,\bar
K^0}(t_4,t_3)|\leq 2\;.
\end{eqnarray}
This has the advantage that it can in principle be tested in
experiments. Alice and Bob insert at a certain distance from the
source (corresponding to the detection times) a piece of matter
forcing the incoming neutral to react. Because the strong
interaction is strangeness conserving one knows from the reaction
products if it is an antikaon or not. Note that different to photons
a $NO$ event does not mean that the incoming kaon is a $K^0$ but
also includes that it could have decayed before. In principle the
strangeness content can also be obtained via decay modes, but Alice
and Bob have no way to force their kaon to decay at a certain time,
the decay mechanism is a spontaneous event. However, a necessary
condition to refute any local realistic theory are \textit{active}
measurements, i.e. exerting the free will of the experimenter (for
more details consult Ref.~\cite{Hiesmayr13}).

Thus the questions is: Can we violate the Bell--CHSH inequality
sensitive to strangeness (\ref{BIstrangeness}) for a certain initial
state and what is the maximum value?

The first naive guess would be yes. In
Refs.~\cite{Hiesmayr4,Hiesmayr13} the authors studied the problem
for an initial maximally entangled Bell state, i.e., $\psi^-\simeq
K^0 \bar K^0-\bar K^0 K^0$, and found that a value greater than $2$
\textbf{cannot} be reached, i.e. one cannot refute any local realistic
theory. The reason is that the particle--antiparticle oscillation is
to slow compared to the decay or vice versa, i.e., the ratio of
oscillation to decay $x=\frac{\Delta m}{\Gamma}$ is about $1$ for
kaons and not $2$ necessary for a violation. On the other hand we have seen that the decay property acts as a kind of ``decoherence''. From decoherence studies we know that some states are more ``robust''
against a certain kind of decoherence than others.

This leads to the question: Exists there another maximally entangled Bell state or another initial non-maximally entangled state which leads to a violation of the above Bell inequality because it is more robust against the decay property?

Indeed this is the case! The violation is not high, it is about $2.1$ (see Ref.~\cite{Hiesmayr13}) and is obtained for a non-maximally entangled state.
This opens for the first time the possibility of a direct experimental test, the remaining question is how to produce any state that violates the Bell inequality.

\subsection{What kind of states do violate the generalized Bell inequality?}

\begin{figure}[t]
  \includegraphics[height=.3\textheight]{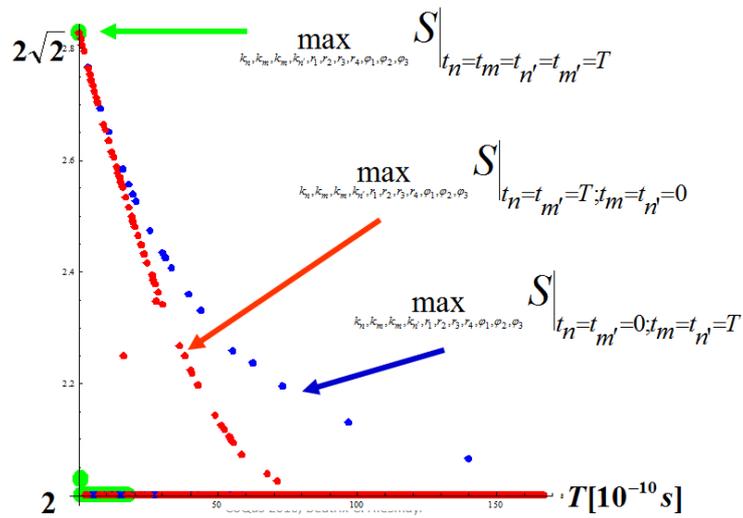}
  \caption{Here the numerically optimized values of the CHSH-Bell type inequality (\ref{chsh}) over all possible quasi-spin choices and any pure state is shown for different choices of times.}\label{figure}
\end{figure}
In principle it is interesting to ask which states are violating the generalized CHSH-Bell type inequality (\ref{chsh}). I.e. we optimize over all four quasispins Alice and Bob can choose in principle and any initial pure state where the four times are given. Thus we have to optimize over $2\times4+6+1=15$ parameters. For that we used a special parametrization and an adapted Nelder-Mead optimization which was also used in optimization of a Bell operator for bipartite qudits (see Refs.~\cite{Spengler1,Spengler2}). Each point in Fig.~\ref{figure} is the maximum over three optimizations, but certainly as it often fails, it has further to be improved. If we choose $t_n=t_m=t_{n'}=t_{m'}=T$ then after a rather short time $T$ no violation of the generalized Bell inequality (\ref{chsh}) can be obtained (big green dots). On the other hand if we choose two times to be equal to zero, $t_n=t_{m'}=T;t_m=t_{n'}=0$, then we find that the violation of the Bell inequality can be obtained for times $T$ which certainly can be realized in experiments (red dots).  If we choose  $t_n=t_{m'}=0;t_m=t_{n'}=T$, the violation is even higher, this reflects the non-symmetry of the measurement choices for the Bell inequality (blue dots).

\section{The kaonic quantum eraser}

Two hundred years ago Thomas Young taught us that photons interfere. Nowadays also experiments
with very massive particles, like the fullerenes, have impressively demonstrated that fundamental
feature of quantum mechanics. It seems that there is no physical reason why not even
heavier particles should interfere except for technical ones. It is well known that the knowledge on the path through the double
slit is the reason why interference is lost. The gedanken experiment of Scully and Dr\"uhl in 1982
\cite{scully82} surprized the physics community, if the knowledge on the path of the particle is
erased, interference is brought back again. Since that work many different types of quantum erasures have been analyzed and experiments were
performed with atom interferometers and entangled photons where the quantum erasure in the so-called
``delayed choice'' mode captures best the essence and the most subtle aspects of the eraser
phenomenon. In this case the meter, the quantum system which carries the mark on the path taken,
is a system spatially separated from the interfering system which is generally called the object
system. The decision \textit{to erase or not} the mark of the meter system ---and therefore
\textit{to observe or not} interference--- can be taken long after the measurement on the object
system has been completed. This was nicely phrased by Aharonov and Zubairy in their review article
\cite{AharonovZubairy} as ``erasing the past and impacting the future''.

Some of these experiments use a kind of double slit device and entanglement. Both are available for neutral kaons, i.e. their time evolution can be view as a double slit varying in time \cite{Hiesmayr10}. In Refs.~\cite{Hiesmayr7,Hiesmayr8,Hiesmayr8a} it was shown that there exist four different types of quantum erasure concepts for neutral kaons:
\begin{itemize}
\item[(a)] Active eraser with \textit{active} measurements
\item[(b)] Partially passive quantum eraser with \textit{active} measurements
\item[(c)] Passive eraser with ``\textit{passive}'' measurements on the meter
\item[(d)] Passive eraser with ``\textit{passive}'' measurements
\end{itemize}
 Two of them $(a),(b)$ can be considered from the theoretical point of view to be analogous to performed erasure experiments with entangled photons. In option $(a)$ the erasure operation is carried out ``actively'', i.e., by exerting the free will of the experimenter, whereas in the
option $(b)$ the erasure operation is carried out ``partially actively'', i.e., the mark of
the meter system is erased or not by a well known probabilistic law, e.g., by a beam splitter or in case of kaons by the decay property.
However, different to photons the kaons can be measured by an \textit{active} or a
\textit{passive} procedure ($=$ decay modes). This offers new quantum erasure possibilities $(c),(d)$ and proves the very
concept of a quantum eraser, namely sorting events.

\section{Summary and discussion of the ``\textit{dynamical nonlocality}'' in the neutral kaons system}

We have shown a generalized Bell inequality (Ref.~\cite{Hiesmayr3}) where both the quasi-spins, i.e. a certain superposition of the strangeness eigenstates (kaon, antikaon), and the times can be varied. Our first choice was to set all times zero and vary in the quasi-spins, by a further trick (Ref.~\cite{Hiesmayr4}), this Bell inequality can be turned into an experimentally testable conditions. In detail we found that no local realistic theory can be constructed which includes the $\mathcal{CP}$--violating property of this system. This is certainly a different feature only available in meson-antimesons systems and nicely relates foundations of quantum mechanics with the very powerful concept of symmetries in particle physics. The disadvantage is that the expectation values involved cannot directly be tested in experiments.

Another choice of measurement settings (quasispins equal the strangeness eigenstates), not lacking this disadvantage, was shown to be non-violating if the initial state is the maximally entangled antisymmetric Bell state. Finding that the decay property (in general of any decaying system) can be modeled by an open quantum formalism gave the idea that there may exist states which are more robust against this ``decoherence'' caused by the decay property. Indeed, such states can be found, in particular a non-maximally entangled state violates the Bell inequality maximally. This opens for the first time the possibility of an experimental test, however, it is still open how to produce such a state and certainly some more subtle problems are awaiting their solution before we will have a conclusive experimental test.

From the theoretical point of view it is interesting to compare the neutral kaon system with other systems. For bipartite qubit systems any pure entangled state violates the Bell inequality and the maximal entanglement is obtained for the maximally entangled states, in contrast to the kaonic two-state system where a non-maximally entangled state maximally violates the Bell inequality. If the entanglement of two spin-$1/2$ particles is considered by an initial observer moving at different speeds, one finds that firstly the entanglement is only Lorentz invariant for certain partitions of the Hilbertspace and secondly that the antisymmetric Bell state behaves differently than the symmetric one  in general (see Ref.~\cite{Friis1}).

 If one considers a generalization of the CHSH-Bell inequality for bipartite entangled qudits, the CGLMP-Bell inequality (Ref.~\cite{Gisin}), one finds for bipartite qutrits that a non-maximally entangled state violates the Bell inequality by a higher amount. Interestingly, no monotonic relation between the amount of entanglement and the violation of the Bell inequality can be found for bipartite qudits, which was demonstrated for the magic simplex states \cite{simplex} in Ref.~\cite{Spengler1}.

Last but not least, we showed that if Alice and Bob could choose any measurement setups and one could produce any initial state for the neutral kaon system, then for rather long times a violation of the Bell inequality can be found.

In summary, the neutral kaon system is a unique laboratory to test fundamental questions of particle physics but as well to test fundamental questions of quantum mechanics.

\vspace{0.5cm}
\textbf{Acknowledgments:}\\
I want to thank Gregor Weihs and Andrei Khrennikov for inviting me to the conference ``Advances in Quantum Theory'' in V\"axj\"o (Sweden) 2010 and to give me the opportunity to organize a session ``Testing foundations in High Energy Physics'' for the forthcoming conference 2011 in V\"axj\"o.

\end{document}